\documentstyle[prl,aps,epsfig,12pt]{revtex}

\begin{document}
 \tolerance 50000

\draft

\title{The Density Matrix Renormalization Group \\ 
applied to single--particle Quantum Mechanics} 

\author{  
M.A. Mart\'{\i}n-Delgado$^{1}$,
G. Sierra$^{2}$ and R.M. Noack$^{3}$
 } 
\address{ 
$^{1}$Departamento de
F\'{\i}sica Te\'orica I, Universidad Complutense. Madrid, Spain.
\\ 
$^{2}$Instituto de Matem\'aticas y F\'{\i}sica Fundamental, C.S.I.C.,
Madrid, Spain. 
\\
$^{3}$Institut de Physique Th\'eorique, Universit\'e de Fribourg,
CH-1700 Fribourg, Switzerland.
}

\maketitle 

\begin{abstract} 
\begin{center}
\parbox{14cm}{ 

A simplified version of White's Density Matrix Renormalization Group (DMRG)
algorithm has been used to find the ground state of the free particle
on a tight--binding lattice.
We generalize this algorithm to treat the tight--binding particle in
an arbitrary potential and to find excited states.
We thereby solve a discretized version of the single--particle
Schr\"odinger equation, which we can then take to the continuum limit.
This allows us to obtain very accurate results for 
the lowest energy levels of the quantum harmonic oscillator,
anharmonic oscillator and double--well potential.
We compare the DMRG results thus obtained with those
achieved by other methods.  
}

\end{center}
\end{abstract}

\pacs{
\hspace{2.5cm} 
}

\section{Introduction}

The Density Matrix Renormalization Group (DMRG) method \cite{DMRG}
originated from the study of a very simple 
problem: the quantum behaviour of a single particle on a lattice
\cite{white-noack}.
The standard renormalization group (RG) method applied to this problem fails 
completely and the understanding of this failure was the key
that led to the DMRG. 
The DMRG has since been sucessfully applied to 
interacting  many--body problems in Condensed Matter and 
Statistical Mechanics and it is very promising in
Atomic, Molecular, Nuclear Physics and Field Theory 
(see Refs.\ \cite{white,dresden-lectures}
for an overall account of the DMRG). 
In view of all these developments, it is interesting 
to come back to the origin of the DMRG and see how it works for 
a single particle under the action of a potential. 
This problem is not purely academic since it provides a testbed for
new ideas and techniques relating to the DMRG in simple models provided by
Quantum Mechanics (QM).
In addition, these simple models can contain interesting physics.
For example, the instantons of gauge theories have a very nice analog
in the double--well potential.
We also hope that the DMRG will provide an accurate new method to 
solve the Schr\"{o}dinger equation numerically.

\section{The problem}

The purpose of this paper is to present a
DMRG algorithm for finding the ground state and
excited states of a particle 
confined to the real line $ -\infty < x <  \infty$, and whose dynamics
is governed  by the Hamiltonian $H= p^2 + V(x)$, where $p^2$ is the 
kinetic energy and $V(x)$ is the potential energy. 
The first step before we can 
apply the DMRG to QM is to discretize the 
Schr\"{o}dinger Hamiltonian $H$.  
This can be done by constraining the position  of the  
particle to take on the discrete values  
$x_n = h (n - \frac{N+1}{2}), (n = 1, 2, \dots, N)$,
where $N$ is the number of allowed sites and 
$h$ is the lattice spacing.
The particle is thus confined to a box of size 
$R= x_N - x_1 =  h (N-1)$. 
After this discretization, the Hamiltonian $H=p^2 + V(x)$ becomes 
the $N \times N$ matrix
\begin{equation}
H_{n,m} = \left\{ \begin{array}{ll}
2/h^2 + V_n & n=m \\
-1/h^2 & |n-m| = 1 \\
0 & {\rm otherwise} \; ,
\end{array} \right.
\label{1}
\end{equation}
\noindent where $V_n = V(x_n)$. Conversely,
one can recover the Schr\"{o}dinger Hamiltonian from Eq.\ (\ref{1}) 
by taking  the  continuum limit $h\rightarrow 0, N \rightarrow \infty$ 
with $R=N\times h$ kept fixed, and then letting $R \rightarrow \infty$.
In the free case, $V(x)=0$, the discrete Hamiltonian,
Eq.\ (\ref{1}), coincides, up to the overall factor $1/h^2$, with the
Hamiltonian studied by White and Noack in Ref.\ \cite{white-noack}. 

\section{The DMRG algorithm}

The problem is to diagonalize the $N \times N$ matrix,
Eq.\ (\ref{1}), using DMRG methods. The basic strategy is to split the 
box of length $N$ into a left block 
$B^L_\ell$ with $\ell$ sites, two middle sites
$\bullet \bullet$ and a right block $B^R_{N- \ell -2}$ with 
$N-\ell-2$ sites, so that the whole system can be represented
as $ B^L_\ell \bullet \bullet B^R_{N - \ell -2}$. If no truncation
is performed, $B^L_\ell$ contains all the degrees of freedom associated
with the $\ell$ sites on the  left hand side, and 
similarly $B^R_{N-\ell-2}$ describes all the degrees of freedom
on the right hand side. 

The DMRG method is based on the idea
that the low--energy properties of the system  
can be described by a few degrees  
of freedom. In particular, if we are only interested
in the ground state of the system we may simply  consider
the blocks $B^L_\ell$ and $B^R_{N - \ell -2}$ to be 
described by a single degree of freedom. 
The effective Hamiltonian of the system then becomes  a $4 \times 4$
matrix whose diagonalization is straightforward. 
Once this is done, the next step is to make a partition of 
the system  either as
$B^L_{\ell+1} \bullet \bullet B^R_{N-\ell-3}$ or as 
$B^L_{\ell-1} \bullet \bullet B^R_{N-\ell-1}$, i.e. make the left or
right side grow by one site, and repeat
the diagonalization. 
One then iterates this procedure, moving the position of the partition,
$\ell$, leftwards and rightwards through the lattice.
After several such sweeps through
the lattice, the DMRG converges to a fixed--point solution
which reproduces the ground state of 
the free tight--binding particle
to high accuracy. 
 
\subsection{Superblock Hamiltonian}

This method can be generalized   
to  treat the
$N_E \geq 1$  lowest energy levels, i.e. $N_E -1$ excited states.
The basic idea is that the left and right blocks
$B^L$ and $B^R$ must contain $N_E$ degrees of freedom,
one for each low energy state. The superblock
Hamiltonian $H_{SB}$ is therefore a $(2 N_E +2) \times (2 N_E +2)$ 
matrix given by
\begin{equation}
H_{SB} = \left( \begin{array}{cccc}
H_L & - v_L & 0 & 0 \\
- v_L^\dagger & h_{CL} & -1/h^2 & 0 \\
0 & -1/h^2 & h_{CR} & -v^\dagger_R \\
0 & 0 & -v_R & H_R \end{array}
\right) \; ,
\label{2}
\end{equation} 
\noindent where $H_L$ and $H_R$ are $N_E \times N_E$ 
matrices, $v_L$ and  $v_R$ are $N_E$--component column vectors,
and $h_{CL}$ and $h_{CR}$ are real numbers. The meaning
of these quantities is as follows: $H_L$ 
is the Hamiltonian which describes the interactions inside
the block $B^L$, $-v_L$ describes the interaction between
$B^L$ and the site next to it in the superblock 
$B^L \bullet \bullet B^R$ and $h_{CL}$ is the Hamiltonian
on a single site. The quantities $H_R, -v_R$ and
$h_{CR}$ for the right half have an analogous meaning.  
The two terms proportional to $-1/h^2$ in Eq.\ (\ref{2}) 
come directly from the off-diagonal terms in Eq.\ (\ref{1}).

The Hamiltonian, Eq.\ (\ref{2}), describes the 
superblock $B^L_\ell \bullet \bullet B^R_{N- \ell -2}$ and
depends on the value of $\ell$, i.e. $H_{SB}= H_{SB}^{(\ell)}$. 
In particular,
the diagonal entries $h_{CL}$ and $h_{CR}$ 
are given by
\begin{equation}
h_{CL} = \frac{2}{h^2} + V_{\ell+1} , \; \;  
h_{CR} = \frac{2}{h^2} + V_{\ell+2} \; .
\label{3}
\end{equation}
Since the blocks $B^{L,R}$ 
contain $N_E$ effective sites, 
$H_{SB}^{(\ell)}$ can be defined for $\ell = N_E, N_E +1, \dots, N- N_E-2$.

\subsection{DMRG truncation}

The superblock can be used either to enlarge the left block
by one site, i.e. $B^L_\ell \bullet \rightarrow B'^L_{\ell+1}$,
or the right block. 
Let us examine how to do this for the left block.
For $N_E=1$, 
the ground state of the superblock Hamiltonian, Eq.\ (\ref{2}),
has four components, which we will designate $(a_L, a_{CL}, a_{CR}, a_R)$.  
The projection of this state onto $B^L \bullet$
yields $(a_L, a_{CL})$, which must then be normalized
by dividing by $N_a = \sqrt{ a_L^2 + a_{CL}^2} $. 
The new effective 
Hamiltonian, $H'_L$, and $v'_L$ are then given by 
\begin{equation}
H'_L = (a'_L, a'_{CL}) \; \left( 
\begin{array} {cc}
H_L & - v_L \\
-v_L & h_{CL} \end{array} \right) \;
\left( \begin{array}{c} a'_L \\ a'_{CL} \end{array} \right), 
\; \; (N_E = 1)
\label{4}
\end{equation}
and
\begin{equation}
v'_L = a'_{CL} \; ,
\label{5}
\end{equation}
where $a'_L = a_L/N_a$ and $a'_{CL}= a_{CL}/N_a$.

We next describe how to generalize this construction to 
$N_E > 1$. 
First we have to obtain 
the lowest $N_E$ eigenstates of $H_{SB}$, 
which we shall denote as $\left\{ ({\bf a}_{L,i}, a_{CL,i}, a_{CR,i}, 
{\bf a}_{R,i}) \right\}_{i=1}^{N_E}$, where ${\bf a}_{L,i}$ and 
${\bf a}_{R,i}$ are $N_E$--component vectors. 
These vectors are projected onto 
a set of $N_E$ vectors of
the block $B^L \bullet$, i.e.,
$\left\{ ({\bf a}_{L,i}, a_{CL,i}) \right\}_{i=1}^{N_E}$. 
Since this set of vectors is not orthogonal in general, we
orthonormalize them explicitly.
Let us call the set of vectors so obtained 
$\left\{ ({\bf a'}_{L,i}, a'_{CL,i}) \right\}_{i=1}^{N_E}$.
The 
projection $B^L \bullet \rightarrow B'^L$ is 
performed by an $(N_E+1) \times N_E$ matrix 
\begin{equation}
A' = \left( \begin{array}{ccc} {\bf a'}_{L,1} & \dots & {\bf a'}_{L,N_E} \\
{a'}_{CL,1} & \dots & {a'}_{CL,N_E} \end{array} \right) \; .
\label{6}
\end{equation}
Operators associated with $B'^L$ are transformed into the new basis via
\begin{equation}
H'_L = A^{\prime \dagger} \; \left( 
\begin{array} {cc}
H_L & - v_L \\
-v_L & h_{CL} \end{array} \right) \; A'
\label{7}
\end{equation}
and
\begin{equation}
v'_{L,i} = a'_{CL,i} , (i= 1, \dots , N_E) 
\label{8}
\end{equation}
Observe that we do not need to construct a density matrix 
to define a unique projection. 
This is a peculiarity of the 
single--particle 
nature of the problem. 
In a many--body problem, we would have to perform the projection using
the reduced density matrix due to the non--single--valuedness of the
projected wave functions.

\subsection{Initialization and sweeps}

The finite--system DMRG algorithm consists of 
a warm-up phase in which 
the system is built up to its actual length $N$,
followed by several leftwards and rightwards ``finite--system'' sweeps
which are repeated until convergence is achieved. 
The initial superblock of the warm-up phase is  taken to be 
$B^L_{N_E} \bullet \bullet B^R_{N_E}$, where
$H_L$  ($H_R$) 
are taken to be the first (last) $N_E$ entries 
of the matrix, Eq.\ (\ref{1}), i.e. 
\begin{eqnarray}
& H_L = ( h_{n,m} , n,m =1, \dots , N_E)  & \label{9} \\
& H_R = ( h_{n,m}, n,m = N-N_E +1, \dots, N) & \nonumber  \; .
\end{eqnarray}
Similarly, $v_L = (0, 0, \dots, 1)$ and 
$v_R = (1,0, \dots, 0)$. For a symmetric potential, i.e.
$V(-x) = V(x)$, the quantities $H_R, v_R$ for the right side 
can be obtained from the left ones by the reflection
operation, $n \rightarrow N_E -n +1$. 
This property applies at each step of the
warm-up phase of the algorithm and corresponds to
the well--known reflection operation of the infinite--system
DMRG algorithm for interacting systems. 
Of course, for a non-symmetric potential
the right block, $B^R$, cannot be formed by reflecting the left 
one, $B^L$.

The behavior of the location of the added site, $\ell$, during the
warm-up process and the first sweep can be summarized 
in the following scheme:
\begin{equation}
\begin{array}{lll}
{\rm warm-up}:    & \ell=& N_E,        \dots,  \frac{N}{2} -1 \\ 
{\rm left}\rightarrow{\rm right}: & \ell=& \frac{N}{2},  \dots,  N- N_E-2 \\
{\rm right}\rightarrow{\rm left}: & \ell=& N- N_E-2,  \dots,  N_E \\
{\rm left}\rightarrow{\rm right}: & \ell=& N_E, \dots, \frac{N}{2}-1 
\end{array}
\; ,
\label{10}
\end{equation}
where we take $N$ to be even. 
We define a sweep to be a complete cycle that ends when the two sites 
of the superblock $B^L_\ell \bullet \bullet B^R_{N -\ell -2}$
are in the middle of the chain.

\section{Analysis of Errors}

In this paper we shall present the results obtained with the
DMRG algorithm applied to three well--known potentials 
and compare them with the results of other methods. 
The three potentials we shall consider are the harmonic oscillator, the
anharmonic oscillator and the double--well potential. 

In order to take the continuum limit, we enlarge
the number of lattice sites, $N$, while taking the discretization step, 
$h$, to be smaller and smaller so that $R=N\times h$.
During this process, it it important to determine the 
accuracy of the DMRG results for the energy levels as $N$ increases.
To this end, we first compare the DMRG results with the known exact
results for the tight--binding particle with fixed boundary conditions
at the ends.
The exact spectrum is given by:
\begin{equation}
  \Psi_n (j) = N_n \sin \frac{\pi (n+1)}{N+1} j,\ \ \ \   
  E_n^{(ex)} =  4 \sin^2(\frac{\pi (n+1)}{2(N+1)})       \label{11}
\end{equation}
where $n = 0,1,\ldots ,N-1$ denotes the energy levels, 
$j = 1,2,\ldots ,N$ are
the lattice sites and the $N_n$ are normalization constants.

\begin{figure}
\begin{center}
\epsfig{width=12cm,file=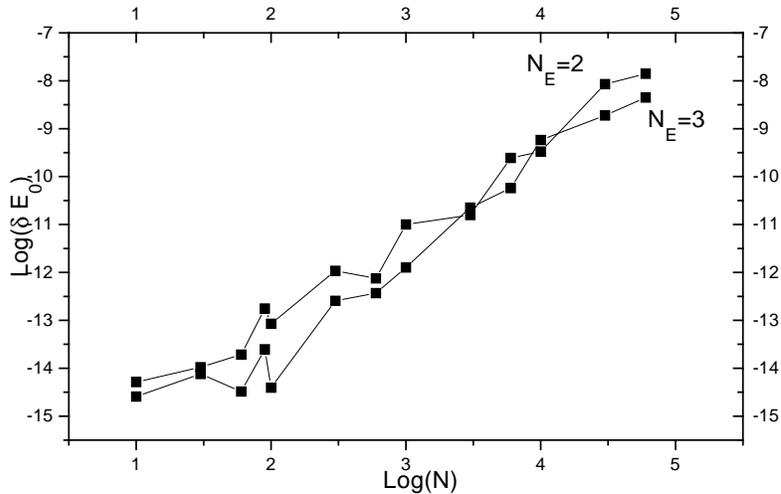}
\end{center}
\caption[]{Logarithm of the relative error in the ground state energy
  of a free particle on a tight--binding lattice as a function of the
  logarithm of the number of sites of the chain, varying the number of
  targeted states, $N_E$.
}
\label{figGSerr} 
\end{figure}

We have made runs targeting the $N_E=1,2,3$ lowest--lying states
and have computed the relative error of the energy levels,
\begin{equation}
\delta E_i = |(E_i(DMRG)-E_i^{(ex)})/E_i^{(ex)}|, \; \; i=1,\ldots,N_E
\; .
\label{12}
\end{equation}
For $N_E=1$, a slowing down of the convergence and, eventually,
numerical instabilities due to roundoff error appear on chains of
length of order $500$ or greater due to a vanishing matrix element
connecting the blocks.
Since this problem does not occur when $N_E \geq 2$, we will take 
$N_E \geq 2$ in the following.

\begin{figure}
\begin{center}
\epsfig{width=12cm,file=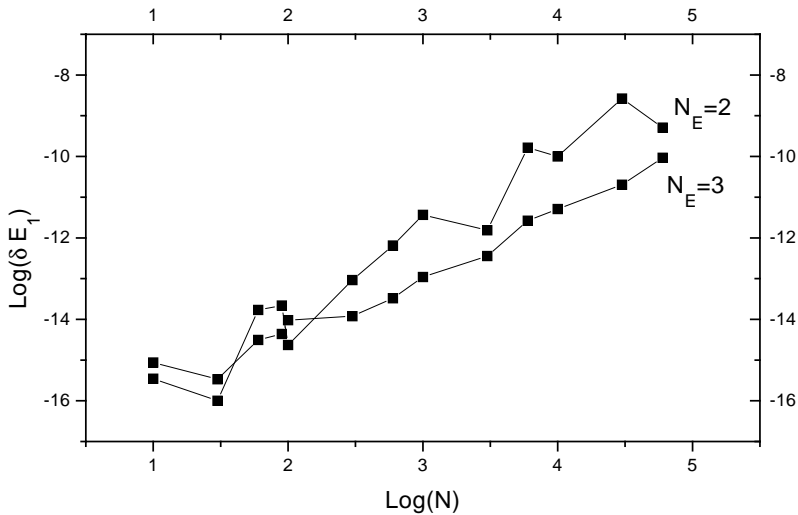}
\end{center}
\caption[]{Logarithm of the relative error in the first excited 
  state energy of a free particle on a tight--binding chain
  as a function of the logarithm of the number of sites of the chain,
  varying the number of targeted states,  $N_E$.
}
\label{figE1err} 
\end{figure}

In Fig.\ \ref{figGSerr}, we plot the relative error in the
ground--state energy for 
$N_E=2$ and $3$ as a function of $N$.
These results were obtained with the finite--system algorithm described
above, taking a sufficient number of finite--system sweeps to obtain
convergence.
Typically, convergence was achieved after five sweeps.
In this plot we see that for a fixed number of targeted states, $N_E$, 
the error increases with the number of sites $N$. 
The range goes from
$\delta E_0 = 5.1 \times 10^{-15}$ for $N=10$ to 
$\delta E_0 = 1.4 \times 10^{-10}$ for $N=60,000$
$(N_E=2)$, 
while $\delta E_0 = 2.6 \times 10^{-15}$ for $N=10$ and 
$\delta E_0 = 4.4 \times 10^{-9}$ for $N=60,000$ $(N_E=3)$. 
Likewise, when the 
number $N_E$ is increased we observe that the errors decrease and this
effect is more apparent when the number of sites $N$ is bigger. 
In Fig.\ \ref{figE1err}, we plot analogous results for the
first excited state, and find that the error behaves essentially the
same as for the ground state.

The lesson to be learned from this analysis is that care must be
taken when going to large lattice sizes because the DMRG tends to
loose numerical accuracy with increasing $N$.
On the other hand, targetting a larger number of states $N_e$
actually {\it improves} the accuracy of each state.
In going to the continuum limit, there is therefore a tradeoff between
the discretization error which can be reduced by taking $N$ large, and
the error in the DMRG results \cite{legeza}.

\section{The harmonic oscillator}

\begin{figure}
  \begin{center}
    \epsfig{width=15cm,file=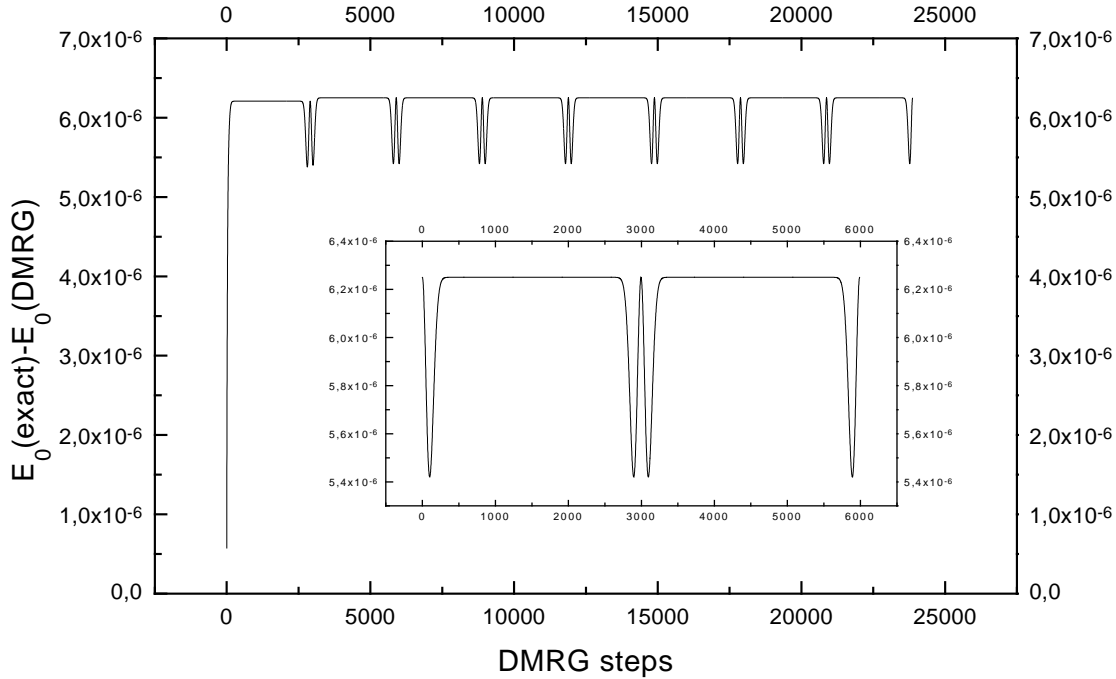}
  \end{center}
  \caption[]{The difference between the ground state energy of the harmonic
    oscillator obtained with 
    the DMRG during 5 sweeps for N=3000 and h=0.01, and the exact continuum
    value. 
    The data plotted starts after convergence in finite--system sweeps
    has been achieved.
    }
  \label{fig3} 
\end{figure}

We first apply the continuum limit of the DMRG to a  
simple and exactly solvable model: the harmonic oscillator. 
We write the Hamiltonian
\begin{equation}
H = P^2 + X^2 \; ,
\label{13}
\end{equation}
where $P=-i {d\over dx}$.
The corresponding exact spectrum with this
normalization is $E_n=2n+1$, $n=0,1,\ldots$~
We can use this example to calibrate the accuracy of the DMRG by
comparing with the exact solution.
In Table\ \ref{table1}, we present the
results for the two lowest eigenstates, for which $E_0=1$ and $E_1=3$. 
These results are obtained using five finite--system sweeps.
We have checked that we obtain the same results to within roundoff
error when performing up to twenty sweeps.
We measure the DMRG energies during the final sweep in the configuration in
which the left block and the right block are the same size, i.e. the
single sites in the superblock are at the middle of the chain.
The error bars in the table are given by the amount of variation in
the energy during the last finite--system sweep, i.e. the value given
is obtained in the last diagonalization step, but the digits in
parentheses vary during previous diagonalizations in the sweep.
We call this method of estimating the error the Global Convergence
Criterium (GCC).
However, the GCC is overly restrictive, as can be seen by examining
the difference between the continuum ground state energy and the DMRG
ground state energy, plotted in
Fig.\ \ref{fig3} as a function of the DMRG step during five
finite--system sweeps for $N=3000$ and $h=0.01$. 
One can clearly see regions in which the solution has been
stabilized as well as depleted regions near the ends of the chains in
which the DMRG energy is higher.
In the inset, we plot the error in energy for a single back-and-forth
sweep on an expanded scale. 
This suggests that the appropriate region to measure the energy must
be located away from the ends. 
We refer to this as a Local Convergence Criterium (LCC).
If we adopt the LCC, 
then the number of stabilized digits increases.
This is something
which we will examine when comparing DMRG results and exact results in the 
forthcoming tables.

In Fig.\ \ref{fig3}, the DMRG ground state energy converges
variationally from above to the exact ground state energy (note that
$E_0(\mbox{DMRG})$ is negated in the plot) for a particular discretization of
the system.
This energy is close to the plateaus in Fig.\ \ref{fig3} and is lower
than the exact continuum energy, i.e.
there is a discretization (or finite--size) correction which is not
necessarily variational, and which is negative here.
This can also be seen in Table\ \ref{table1} for the exact
diagonalization as well as for the DMRG results.

\begin{table}
\begin{center} 
\begin{tabular}{|c|c|c|c|c|c|} 
\mbox{Method} & $N$ & $h$ & $R =N\times h\ $ & $1.0 - E_0(N)$ & $3.0 - E_1(N)$ \\  \hline \hline

\mbox{DMRG}        & 100 & 0.1 & 10 & $6.253912540(370)\times 10^{-4}$ & $3.128521663(378)\times 10^{-3}$ 
\\ \hline

\mbox{Exact Diag.} & 100 & 0.1 & 10 & $6.253912540\;310\;\times 10^{-4}$ & $3.128521663\;376\;\times10^{-3}$ 
\\ \hline

\mbox{DMRG}        & 1,000 & 0.01 & 10 & $6.24989(322)\times 10^{-6}$ & $3.1243338(564)\times 10^{-5}$ 
\\ \hline

\mbox{Exact Diag.} & 1,000 & 0.01 & 10 & $6.24989\;256\;\times 10^{-6}$ & $3.1243337\;792\;\times 10^{-5}$
\\ \hline
  
\mbox{DMRG} & 5,000  & 0.002 & 10 &   $2.498(36)\times 10^{-7}$ & $1.2427(28)\times 10^{-6}$
 \\ \hline   

\mbox{DMRG} & 10,000 &  0.001 & 10  &  $6.2(45)\times 10^{-8}$ & $3.051(90)\times 10^{-7}$
\\ \hline
 
\mbox{DMRG} & 20,000 & 0.0005 & 10 & $1.(56)\times 10^{-8}$ & $7.0(80)\times 10^{-8}$ 
\\ \hline

\mbox{DMRG} & 20,000 & 0.001 & 20  & $6.2(64)\times 10^{-8}$ & $3.12(50)\times 10^{-7}$
\\ \hline

\mbox{DMRG} & 50,000 & 0.0002 & 10  & $(2.87)\times 10^{-9}$ & $(4.99)\times 10^{-9}$ 
\\ \hline

\mbox{DMRG} & 50,000 & 0.0004 & 20 & $(0.961)\times 10^{-8}$ & $(5.00)\times 10^{-8}$ 
\\ \hline

\mbox{DMRG} & 100,000 & 0.0001 & 10 & $(-0.71)\times 10^{-8}$ & $(-4.15)\times 10^{-8}$ 
\\ \hline

\mbox{DMRG} & 100,000 & 0.0002 & 20 & $(0.16)\times 10^{-8}$ & $1(.24)\times 10^{-8}$\\ 
\end{tabular}
\end{center}
\caption{The difference between the exact continuum energies, $E_n = 2n+1$, 
and the energies on the discretized lattice, $E_0(N)$ and $E_1(N)$, for 
the ground and first excited states of the harmonic oscillator 
obtained with the DMRG and exact diagonalization.
Here $N$ is the number of sites on the chain, $h$ is the discretization
step and $R$ is the size of the continuum system.
The errors (digits in parentheses) are determined using the GCC procedure 
explained in text.}
\label{table1}
\end{table}

In Table\ \ref{table1} we present several runs in which the final
size $R$ of the continuum system is varied in order to check that a
convergence has been achieved to a certain number of digits.
If $R$ is large enough, the effect of changing it is not very large
due to the exponential fall-off of the wave functions. 
We find similar dependence of the error on $R$ for higher excited states.
Thus we find that the exact spectrum is very well reproduced by the
continuum limit of the DMRG.

In Table\ \ref{table1} we also show, when possible, energies obtained
via exact diagonalization of the Hamiltonian matrix, Eq.\ (\ref{1}).
This is possible for system sizes of up to about $N=2000$.
Exact diagonalization is much more expensive in computer time than
DMRG, and yields energies for a particular system size that agree with
the DMRG to 14 significant digits.
This level of agreement is quite remarkable given that the largest
matrix diagonalized in the DMRG procedure is $6\times6$.

\section{The anharmonic oscillator}

Now that we have checked that the continuum limit of the DMRG
accurately reproduces a well-known solvable case, we apply DMRG to the
anharmonic oscillator, which has no closed analytical solution
(see \cite{galindo} and references therein). 
Here our purpose is to compare the efficiency of
DMRG to that of other standard methods employed in single--particle
quantum mechanics.
We treat the Hamiltonian
\begin{equation}
H = P^2 + X^2 + C X^4 \; ,
\label{14}
\end{equation}
where $C$ is a positive coupling constant and we have normalized the
mass term.

In Table\ \ref{table2}, we display the DMRG results obtained using a
similar analysis as for the harmonic oscillator. 
We compute the two lowest energy states, compare them with Exact
Diagonalization results, and obtain the same accuracy as for the
harmonic oscillator. 
We also compare with the results of a number of
methods commonly used in standard quantum mechanics such as the Hill
determinant method \cite{hill}, Borel-Pad\'e approximants 
of the perturbation series \cite{borel} and variational computations
in the energy basis of the harmonic oscillator \cite{zara}. 

We have computed the state energies for various coupling constants
$C=0.1, 1, 10$, ranging from weak-- to strong--coupling.
We notice that while the Hill determinant method
and the Borel-Pad\'e perform better for weak couplings, the continuum
limit of the DMRG performs equally well for the whole range of
couplings. 
This is due to the variational
nature of the method.

\begin{table}
\begin{center} 
\begin{tabular}{|c|c|c|c|c|c|c|} 
\mbox{Method} & $N$ & $h$ &  
$R =N\times h$ & $E_0$ & $E_1$ & $C$~ \\  \hline \hline
\mbox{DMRG}& 30,000 & 0.001 & 30 & 1.06528543(310) & 3.30687158(195) & 0.1~~ \\ \hline 
\mbox{DMRG}& 20,000 & 0.001 & 20 & 1.06528543(289) & 3.30687158(195) & 0.1~~ \\ \hline 
\mbox{Borel-Pad\'e \cite{borel}} & & & & 1.06528550 954~ & & 0.1~~ \\ \hline
\mbox{Hill meth. \cite{hill}}    & & & & 1.06528550 954~ & & 0.1~~ \\ \hline

\mbox{DMRG}& 30,000 &  0.001 &  30 &  1.39235148(046) &  4.64881163(809)  & 1~ \\ \hline 
\mbox{DMRG}& 20,000 &  0.001 &  20 &  1.39235148(038)  & 4.64881163(809)  & 1~ \\ \hline 
\mbox{Borel-Pad\'e \cite{borel} }& & & & 1.392350(6)~~~~~~ & & 1~ \\ \hline
\mbox{Hill meth. \cite{hill}} & & & & 1.3925316~~~~~~~~ & & 1 \\ \hline

\mbox{DMRG}& 30,000 & 0.001 & 30 & 2.449173(484) & 8.5989993(093) & 10~ \\ \hline 
\mbox{DMRG}& 20,000 & 0.001 & 20 & 2.449173(484) & 8.5989993(093) & 10~ \\ \hline
\mbox{Variational \cite{zara}}& & & & 2.449174 072~ & & 10~ \\ \hline
\mbox{Borel-Pad\'e \cite{borel} } & & & & 2.440(527)~~~~ & & 10~ \\ \hline
\mbox{Hill meth. \cite{hill}} & & & & 2.4491740~~~~~ & & 10~ \\
\end{tabular}
\end{center}
\caption{DMRG results for the Anharmonic Oscillator 
$H=P^2 + X^2 + C X^4$ in the continuum limit. 
The notation is the same as that in Table\ \protect\ref{table1}. 
}
\label{table2}
\end{table}

From Table\ \ref{table2}, one can see that the agreement of the DMRG
with the other methods is excellent. 
Moreover, once the DMRG results have converged, 
the results are lower in energy than those of the other methods. 
Thus, since DMRG is variational,
we conclude that the DMRG results are closer
to the exact results than those of the other methods.

\section{The double-well potential}

We hope that it is now clear that the DMRG method is an excellent method to
compute the energy spectrum and wave functions in quantum
mechanical problems. 
We will now apply it to an anharmonic oscillator with a potential 
in the shape of a double well. 
This problem is particularly interesting for several reasons.
One, as we shall see, is that the system has a  tunable gap which can
be used to investigate the dependence of the convergence of the DMRG
on the size of the gap, 
and the other is its potential as a new
non-perturbative method for Quantum Field Theory problems.
We shall not pursue the latter goal here but will only point out
that a number of non--perturbative Renormalization Group techniques
have been established since 
Wilson's original formulation of the RG \cite{wilson-kogut}. 
These formulations are called Exact Renormalization Group 
(ERG) because they are based in exact RG flow equations 
(see \cite{faro} for a recent review on this subject). 
However, since these exact equations are
not exactly solvable in general, one usually has to resort to
approximate methods in practical applications.
The question then arises as to how good these approximations are. 
In order to check their validity,
they are applied to well-known problems in single--particle Quantum
Mechanics. 
The rationale is that if they are not even able to quantitatively
reproduce the physics of these simple systems,
their application to truly field theoretical problems would be even less
successful. 

The Hamiltonian for this potential reads
\begin{equation}
H = P^2 - X^2 + C X^4 \; ,
\label{15}
\end{equation}
where $C$ is the coupling constant and we have normalized the negative
mass term.
The potential has two minima at the positions
\begin{equation}
x_{0}^{(\pm)}=\pm {1\over \sqrt{2 C}} \; .
\label{16}
\end{equation}
Classically, these minima are degenerate in energy. 
If treated perturbatively, quantum fluctuations can modify the
classical energy but not lift the degeneracy. 
Splitting of the energy levels can only
occur if quantum tunneling between the two wells is taken into account.

In this symmetric potential, the energies can be arranged
into pairs $E_n^{(\pm)}$ depending on their
even $(+)$ or odd $(-)$ parity.
The energy gap from the ground state to the first excited state is then
defined as
\begin{equation}
\Delta_0(C) =  E_0^{(-)}(C) - E_0^{(+)}(C) \; .
\label{17}
\end{equation}

In the weak coupling limit ($C$ very small), the gap
$\Delta_0(C)$ can be computed using an instanton approximation plus
higher corrections \cite{zinn-justin}, giving the asymptotic
formula
\begin{equation}
\Delta_0(C) = 8 \sqrt{{\sqrt{2}\over \pi C}} e^{-{\sqrt{2}\over 3C}} 
[ 1 - {71\over 1!} ({2\sqrt{2}\over 12}) C -  {6299\over 2!} ({2\sqrt{2}\over 12})^2 C^2 
 - {2691107\over 3!} ({2\sqrt{2}\over 12})^3 C^3 + O(C^4)]
\label{18}
\end{equation}
This exponentially decreasing behavior produces an essential
singularity in the energy gap.
Our purpose is to capture this highly non-trivial 
behavior with the DMRG.

One can understand what happens to the system in
the weak coupling limit on physical grounds.
From Eq.\ (\ref{16}), we see that the distance between the minima diverges
as $C \rightarrow 0$, so that we effectively end up with a system
formed by two independent potential wells. 
The system then becomes exponentially degenerate for
each pair of energy levels and thus gapless.

It is also interesting to use this quantum mechanical example
to test the behavior of DMRG for gapless systems. 
This is a very important issue when dealing with the
strongly interacting quantum many-body systems to which DMRG is
usually applied.
Since its early development, it has been known that the
the DMRG method \cite{DMRG} produces much more accurate
results for finite correlated systems with a gap than for gapless
systems.
For gapless systems one has to use the finite--system algorithm 
on larger system sizes than for gapful systems in order to obtain 
results of comparable accuracy \cite{ABO}.
We shall show below that the DMRG handles well the case in which the gap
between the ground state and the first excited state becomes
negligibly small, i.e. when the two minima of double well
potential are far apart.

\begin{table}
\begin{center} 
\begin{tabular}{|c|c|c|c|c|c|} 
\mbox{Method} & $C$ & $N$ & $h$ & $E_0$  & $\Delta$\\  \hline \hline
\mbox{DMRG}  & 1 & 1000 &   0.01      &   
0.65764425361(29)  & 2.176825710(298) \\ \hline 
\mbox{Exact Diag.} & 1 & 1000  &  0.01 &   
0.6576442536117 &  2.176825710302 \\ \hline

\mbox{DMRG} & 1 & 20,000 & 0.0005 &      
0.657652983(568)  & 2.17688305(297) \\ \hline 
\mbox{R-R}  & 1 & & & 
0.657653005181~~ &  2.17688319705~~ \\  \hline

\mbox{DMRG} & 0.6 & 20,000 & 0.0005 &    
0.39195261(873)  & 1.6332847(928) \\ \hline
\mbox{R-R}  & 0.6 & & &  
0.39195263337~~  & 1.6332848846~~ \\ \hline

\mbox{DMRG} & 0.1 & 40,000 & 0.001 & 
-1.26549292(138)         & 0.11243368(739) \\ \hline
\mbox{R-R}  & 0.1 & & & 
-1.26549283721~~         & 0.11243370614~~~\\ \hline
\mbox{Inst.} & 0.1 & & & & 0.11447450849~~ \\ \hline

\mbox{DMRG} & 0.06 & 60,000 & 0.0005 &      
-2.82363949(203)          & 0.0072997661(673) \\ \hline
\mbox{R-R} & 0.06 & & &  
-2.82363945845~~          & 0.0072997526870~~ \\ \hline
\mbox{Inst.} & 0.06 & & & & 0.0073139070463~~ \\ \hline

\mbox{DMRG} & 0.02 & 90,000 &  0.0004 & 
-11.106472434(074) & $2.1(074)\times 10^{-9}$ \\ \hline
\mbox{R-R} & 0.02 & & &  
-11.106472414954~~ & $2.1043~~\times 10^{-9}$ \\ \hline
\mbox{Inst.} & 0.02 & & & & $2.10737\times 10^{-9}$  
\\  \hline
\end{tabular}
\end{center}
\caption{DMRG results for the double-well potential 
$H=P^2 - X^2 + C X^4$ in the continuum limit. 
The notation is the same as that in Table\ \ref{table1}. 
R-R stands for the results obtained with the Rayleigh-Ritz method 
explained in the text.}
\label{table3}
\end{table}

In Table\ \ref{table3}, we present the results for the lowest two energy
levels and the gap upon varying the coupling constant from $C=1.0$ to
$C=0.1$. 
These values range from the strong-- to the weak--coupling regime. 
The DMRG results are obtained as described previously after five
finite--system sweeps.
We also compare with Exact Diagonalization methods when $N < 2000$ and 
find excellent agreement, typically up to 10 
digits or better.

\begin{figure}
  \begin{center}
    \epsfig{width=14cm,file=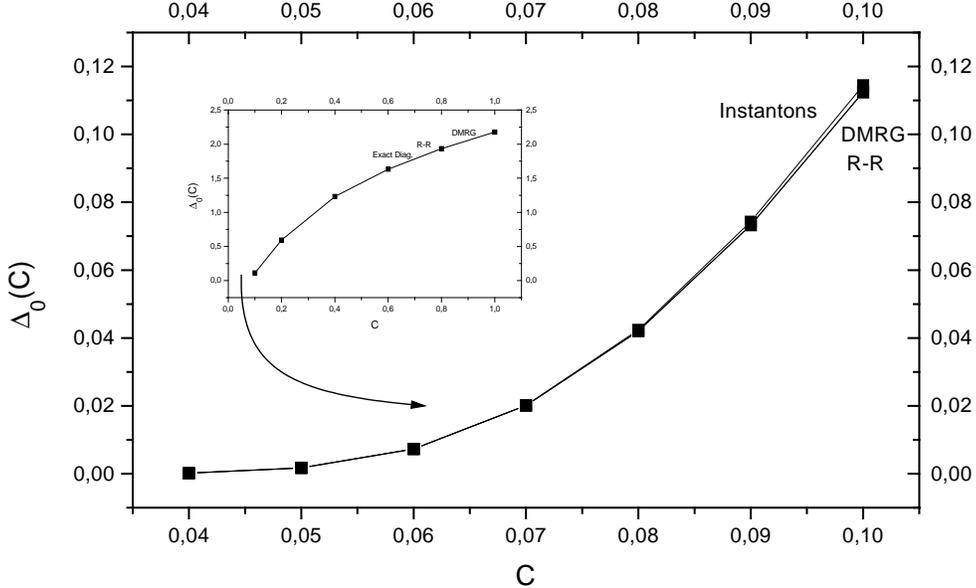}
  \end{center}
\caption[]{Energy gap $\Delta_0(C)$ in the anharmonic double-well
  potential as a function of the coupling constant $C$ in the very
  weak coupling regime (Table\ \protect\ref{table3}).  
  Here the results of the three methods: the DMRG, 
  Instanton calculation and Rayleigh--Ritz (R--R) overlap. 
  Results for a wider range of $C$ (Table\ \protect\ref{table3}) are
  shown in the inset.}
\label{fig4} 
\end{figure}

In Table\ \ref{table3}, we have also computed the energy variationally
using 
the Rayleigh-Ritz method \cite{galindo} (R--R). 
We compute the expectation value of the Hamiltonian, Eq.\ (\ref{15}), in
the energy basis of the harmonic oscillator consisting of up to
$M=1000$ states.
In this representation, the non-vanishing elements of the Hamiltonian
lie within a band and are given by
\begin{eqnarray}
\langle n|H| n\rangle & = & C [{1\over 4} n(n-1) + {1\over 4} (2n+1)^2 + {1\over 4} (n+1)(n+2)] \nonumber \\
\langle n|H| n+2\rangle & = & -\sqrt{(n+2)(n+1)} + C [\sqrt{(n+1)(n+2)}(n+{3\over 2})] \nonumber \\
\langle n|H| n+4\rangle & = &  C {1\over 4} \sqrt{(n+4)(n+3)(n+2)(n+1)} \; .
\label{19}
\end{eqnarray}
In this representation, the Hamiltonian is already in the continuum
limit.
The Rayleigh-Ritz theorem states that the $M$ resulting
energy levels will be upper bounds to the first $M$ exact energy levels. 
As seen in Fig.\ \ref{fig4}, the agreement between exact and DMRG methods is
excellent for this range of the coupling constant and, in fact, the
curves appear overlapped in the plot.

Recently, one ERG method has been applied to the study of the
double-well potential with the aim of probing the method in the whole
range of coupling constants \cite{hori}. 
This method is based on the solution of the local potential
approximation of the Wegner-Houghton equation \cite{wegner,ka}. 
The outcome of these investigations is that the ERG
performs very well in the strong coupling regime where no
perturbation treatment is available.
However, in the weak coupling limit, the ERG fails to reproduce the
behavior found with the instanton formula, Eq.\ (\ref{18}).

To check the performance of the DMRG as compared to the instanton
formula, Eq.\ (\ref{18}), in the very weak
coupling limit, we have extended our computations from $C=0.1$ down to
$C=0.02$, where the gap becomes as small ${\cal O}(10^{-9})$. 
In Table\ \ref{table3} and in Fig.\ \ref{fig4}, we present the results
from the DMRG, R-R and the instanton formula. 
We see that DMRG is an excellent
non-perturbative method in the entire range of coupling constants and
that it is able to quantitatively capture the exponentially decreasing
behavior of the gap.

\section{Conclusions and Prospects}

The DMRG method we have presented in this paper
is a natural extension of one introduced
by White to illustrate the DMRG algorithm for the most simple
problem: a free particle on a tight--binding chain \cite{white}. 
We have shown that the DMRG works with 
very high precision, yielding  
an accurate determination of the lowest energy levels
for three different potentials: the harmonic oscillator,
the anharmonic oscillator and the double--well potential.
Its performance is better than or comparable to other known
perturbative and non--perturbative methods. 
For single--particle Quantum Mechanics, the DMRG
does not require the use of a density matrix. 
The number of states retained in the RG procedure is equal
to the number of states to be obtained, $N_E$.
Aside from $N_E$, the variational wave function of the DMRG 
has no adjustable parameters.

Single--particle Quantum Mechanics has been widely used in the past as
a testing ground for concepts or techniques that can be applied to more
complex systems. 
With this in mind, we have studied quantum tunneling
through a potential barrier for the double-well potential
and found a value of the gap very close to the exact one for a large
range of coupling constants. 
This is an interesting result
because it shows that the DMRG can, in principle, handle tunneling
phenomena better than other methods such  as the Exact Renormalization 
Group \cite{hori}. 
An interesting topic for future work would be to explore to what
extent this feature holds for many--body systems or for field theory. 

Although the DMRG was originally developed as a ground state
technique, there have recently been new developments in using it to
obtain dynamical information \cite{Karen,Kuhner}.
In the context of single--particle Quantum Mechanics, one could
ask whether the DMRG could give information 
about  phase shifts, decay rates, etc. 
These and other questions remain to be investigated. 


\section*{Acknowledgements} We would like to thank 
 A.\ Galindo and S.R.\ White for useful conversations on this topic.

We would like to thank 
the Max Planck Institute for the Physics of Complex Systems in
Dresden for support to participate in
the DMRG98 Seminar/Workshop, at which the present 
work was initiated.

M.A.M.D. and G.S. acknowledge support from the 
DIGICYT under contract No. PB96/0906, and R.M.N acknowledges support
from the Swiss National Foundation under Grant Nos. 20--46918.96 and
20--53800.98.


\end{document}